\documentclass[letterpaper,twocolumn,showpacs,pra,aps]{revtex4-1}

\usepackage{amsmath}  
\usepackage{amsfonts} 
\usepackage{graphicx} 
\usepackage{amssymb}

\begin{document}

\title{Normal modes of vibrations around Hubble flow in Jellium}

\author{Eugene B. Kolomeisky}

\affiliation
{Department of Physics, University of Virginia, P. O. Box 400714, Charlottesville, Virginia 22904-4714, USA}

\date{\today}

\begin{abstract}
A macroscopic Coulomb system of identical charged particles with or without a compensating background charge can evolve maintaining spatial homogeneity and isotropy that mimic the cosmological evolution of a universe with repulsive gravity.   Here we study dynamics of small perturbations on the background of the corresponding Hubble flow by analyzing its normal modes of vibrations.  Arbitrary disturbance of the flow can be resolved into two electro-acoustic, two vortical, and one entropic modes whose dynamics is investigated.   Specifically, in the zero pressure or long-wavelength limits perturbations of density and velocity evolve in a manner that is independent of the form of the initial disturbance.  The same conclusion applies to vortical perturbations of the velocity for arbitrary pressure while entropic perturbations are advected by the Hubble flow.   Without the background charge the underlying Hubble flow describes a Coulomb explosion whose stability with respect to small disturbances is also demonstrated.      
\end{abstract}


\maketitle

\section{Introduction}

Jellium -- a one-component plasma of interacting electrons on the background of a uniform positive charge (representing the effect of ions) \cite{Wigner}-- is a paradigm of physics.  The notion first appeared in Thomson's static plum pudding model of the atom \cite{Thomson}, in which the electrons are immersed inside a positively charged uniform ball representing the nucleus.    While Rutherford's experiments ruled out this model as describing the atom, it nevertheless continued to offer valuable insights beyond its original purpose.  For example, in condensed matter systems, Jellium has shaped our understanding of conductors \cite{Bohm_Pines,Pines_Nozieres,Mahan}.  The model is also a starting point in describing artificial atoms - systems of excess electrons confined in semiconductor quantum dots - where confinement potential plays a role of compensating background charge \cite{Bednarek}.  For example, molecular dynamics simulations provide evidence that the lowest energy configuration in the interior of a finite Coulomb cluster changes from a shell structure to a bcc lattice as the number of particles increases \cite{Totsu}.  The Jellium model also successfully describes the interior of white dwarfs \cite{Salpeter} where the roles of the electrons and ions are reversed:  nuclei move in a uniform electron gas.  The classical Thomson model is additionally an inspiration to the surface Coulomb problem \cite{surface} whose goal is to determine lowest energy configurations of identical point charges on the surface of a sphere. 

Recent analysis \cite{EBK} has demonstrated that Jellium can also flow in a homogeneous and isotropic manner according to equations that have the structure of the cosmological equations of the general theory of relativity \cite{ZN, Peebles,Mukhanov}.  There is a Hubble law, and the background charge (if present) mimics the effect of a negative cosmological constant.  Specifically, evolutions without the background charge describing Coulomb explosions imitate the non-singular open cosmologies in negatively curved spaces, while breathing modes in conductors model oscillatory universes including the anti-de Sitter space.  The analogy between Coulomb explosion and Hubble flow was previously mentioned in Ref.\cite{Kaplan}.

The relationship between the flow of Jellium and cosmological evolutions is antipodal:  while gravity is attractive, Jellium is made of repulsive particles which leads to qualitatively different physical properties of the two systems.  An example relevant to the present study is the problem of the gravitational instability which is the root cause behind the formation of galaxy clusters, galaxies and stars \cite{ZN,Peebles,Mukhanov}.  The goal of this work is to study the Jellium counterpart to the problem of gravitational instability.  While it is clear from the outset that repulsive Coulomb interactions do not promote an instability of the underlying Hubble flow, there are two primary reasons why the problem is relevant:  

First of all, it is important to find out whether the Jellium-cosmology mapping \cite{EBK} goes beyond homogeneous and isotropic evolutions.  In a nutshell, the answer is affirmative, but the repulsive character of Coulomb interactions leads to qualitatively different physics consequences.

Secondly, strict Hubble flows of Jellium require homogeneous and isotropic initial conditions as well as lack of disturbances in the course of evolution which in laboratory experiments are challenging to avoid.  Understanding how inhomogeneities evolve is important for the interpretation of experimental observations. 

Past theoretical studies of Coulomb explosion \cite{Kaplan,Kovalev05,Kovalev07,Novikov08,Andri} have been motivated by applications in the context of interaction of high power lasers with nano-sized targets.  Their focus has been finite-size and spherically symmetric systems with special emphasis on conditions underlying formation of shock waves as well as effects of surfaces of finite non-neutral plasma on density perturbations \cite{Saalman}.  On the hand, the results presented in this work may, in principle, be applied to extended plasmas (both neutral and non-neutral).  While the theory given here is linear, no restrictions are made on the type of perturbation of Hubble flows. 

\section{Statement of the problem}                      

Our analysis is based on a classical macroscopic theory that combines hydrodynamics and electrostatics \cite{Feynman}.  In this approach Jellium is  treated as an ideal charged liquid characterized by the local position- and time-dependent number density $n(\textbf{r},t)$ and velocity $\textbf{v}(\textbf{r},t)$ fields.  These are related by the continuity equation
\begin{equation}
\label{continuity}
\frac{\partial n}{\partial t}+\nabla \cdot(n\textbf{v})=0.
\end{equation}
The equation of motion of the liquid is given by the Euler equation of hydrodynamics 
\begin{equation}
\label{2nd_law}
\frac{\partial \textbf{v}}{\partial t} +(\textbf{v}\cdot \nabla)\textbf{v}=- \frac{e}{m}\nabla \varphi-\frac{1}{mn}\nabla P
\end{equation}
where $m$ is the particle mass, $e$ is its charge, $\varphi$ is the electrostatic potential, and $P(n,s)$ is the pressure which is a function of the number density $n$ and the entropy per unit mass $s$.  The latter obeys the equation
\begin{equation}
\label{entropy_continuity}
\frac{\partial s}{\partial t}+\textbf{v} \cdot \nabla s=0
\end{equation}
expressing the adiabatic character of the motion.  The charged liquid is accelerated both by the bulk electric force, the first term in the right-hand side in Eq.(\ref{2nd_law}), and by the gradient of the pressure $\nabla P$.  The electrostatic potential $\varphi$ in turn is determined by the Poisson equation
\begin{equation}
\label{general_Poisson}
\nabla^{2}{\varphi}=-4\pi e(n-n_{0}),
\end{equation}
where $n_{0}$ is the number density of the oppositely charged background.  It is assumed that the motion is non-relativistic and thus magnetic effects are neglected. 

We finally note that the validity of the continuum description has been recently verified by comparing conclusions of Newtonian cosmology with expansion dynamics of a system of point masses \cite{Fleury}.  There is no doubt that hydrodynamic approach adopted in this paper to investigate the Coulomb counterpart of the problem is also an adequate description. 

\subsection{Summary of Hubble evolutions in Jellium}

The system of equations (\ref{continuity})-(\ref{general_Poisson}) admits a class of exact spatially homogeneous and isotropic evolutionary solutions with the following properties \cite{EBK}:

\begin{enumerate} 
\item Relative velocity $\textbf{v}$ of any two particles of the liquid and their separation vector $\textbf{r}$ are related by Hubble's law
\begin{equation}
\label{Hubble_law}
\textbf{v}=H(t)\textbf{r}
\end{equation}
where $H(t)$ is the Hubble parameter.  Hubble's law can be written in an equivalent form
\begin{equation}
\label{comoving}
\textbf{r}=a(t)\textbf{x}
\end{equation}
where $\textbf{x}$ is a time-independent relative comoving coordinate vector defining the particle pair considered.  The universal function $a(t)$, the scale factor, is defined in terms of the Hubble parameter as
\begin{equation}
\label{scale_factor}
H(t)=\frac{\dot{a}}{a}
\end{equation}
where the dot is a shorthand for the derivative with respect to time.  

\item For an observer at $\textbf{r}=0$ the electrostatic potential at position $\textbf{r}$ is given by 
\begin{equation}
\label{background_potential}
\varphi=-\frac{2\pi e}{3}[n(t)-n_{0}]r^{2}.
\end{equation}
The evolving density $n(t)$ and Hubble's parameter $H(t)$ are related by the two equations
\begin{equation}
\label{Hubble_definition}
\dot{n}=-3H(t)n,
\end{equation}
\begin{equation}
\label{Friedmann1}
\dot{H}+H^{2}(t)=\frac{4\pi e^{2}}{3m}(n-n_{0}).
\end{equation}
Eqs. (\ref{Hubble_definition}) and (\ref{Friedmann1}) combined with the definition of the scale factor $a(t)$ (\ref{scale_factor}) can be integrated resulting in relationships
\begin{equation}
\label{charge_conservation}
n(t)=\frac{\beta}{a^{3}(t)}
\end{equation}
\begin{eqnarray}
\label{energy_integral0}
\frac{m\dot{a}^{2}}{2}&+&U(a)=E,\nonumber\\
U(a)&=&\frac{4\pi e^{2}}{3}\left (\frac{\beta}{a}+\frac{n_{0}a^{2}}{2}\right )
\end{eqnarray}
where $\beta>0$ and $E$ are integration constants.  Eq.(\ref{energy_integral0}) can be integrated resulting in a $t(a)$ dependence
\begin{equation}
\label{quadrature}
t=\sqrt{\frac{m}{2}}\int^{a}\frac{da}{\sqrt{E-U(a)}}.
\end{equation}

\item The character of the flow can be visualized by viewing Eq.(\ref{energy_integral0}) as a statement of conservation of energy for a particle of energy $E$ and position  $a\geqslant 0$ moving in the field of the potential energy $U(a)$.  For $n_{0}$ finite $U(a)$ is a potential well with a minimum at $a=a_{0}=(\beta/n_{0})^{1/3}$ which according to Eq.(\ref{charge_conservation}) corresponds to the state of local neutrality $n=n_{0}$.  In the vicinity of $a=a_{0}$, the potential energy function can be approximated as
\begin{equation}
\label{parabolic_approximation}
U(a)= 2\pi e^{2}(n_{0}\beta^{2})^{1/3}+\frac{m\omega_{p}^{2}(a-a_{0})^{2}}{2}
\end{equation} 
where $\omega_{p}$ is the plasma frequency defined as \cite{Feynman}
\begin{equation}
\label{plasma_frequency}
\omega_{p}^{2}=\frac{4\pi n_{0}e^{2}}{m}.
\end{equation}
The implication is that Eq.(\ref{energy_integral0}) has solutions if $E\geqslant 2\pi e^{2}(n_{0}\beta^{2})^{1/3}$, and that for $E=2\pi e^{2}(n_{0}\beta^{2})^{1/3}+0$ the flow has the character of a harmonic breathing oscillation with the plasma frequency (\ref{plasma_frequency}).  

\item For $E> 2\pi e^{2}(n_{0}\beta^{2})^{1/3}$ the Hubble flow is an anharmonic breathing oscillation limited by the two turning points $a_{1}$ and $a_{2}\geqslant a_{1}$, solutions to the equation $U(a_{1,2})=E$, corresponding to the largest and the smallest densities of Jellium, respectively.  A notable special case of this regime that mimics the anti-de Sitter space corresponds to the $E\gg2\pi e^{2}(n_{0}\beta^{2})^{1/3}$ condition.  Then $a_{1}\rightarrow 0$ and 
\begin{equation}
\label{anti}
a(t)=a_{2}\left |\sin\frac{\omega_{p}t}{\sqrt{3}}\right |, ~~~~~a_{2}=\left (\frac{6E}{m\omega_{p}^{2}}\right )^{1/2}
\end{equation}
which is a series of the frequency $2\omega_{p}/\sqrt{3}$ sinusoidal bounces.  Eq.(\ref{anti}) applies provided
\begin{equation}
\label{constraint}
a(t)\gg a_{1}=\frac{4\pi e^{2}\beta}{3E}.
\end{equation}

\item In the absence of the background charge, $n_{0}=0$, the motion described by Eq.(\ref{energy_integral0}) is infinite, $a_{2}=\infty$, and the Hubble flow corresponds to a Coulomb explosion:
\begin{eqnarray}
\label{explosion_flow}
a&=&\frac{a_{1}}{2}(\cosh\xi+1),\nonumber\\
t&=&\frac{1}{4}\left (\frac{3ma_{1}^{3}}{2\pi e^{2}\beta}\right )^{1/2}(\sinh\xi+\xi)
\end{eqnarray}  
where $\xi$ is a parameter varying between minus and plus infinity;  at $\xi=0$ Jellium has its largest density.  This is when a contraction or implosion ($t<0$) changes to an expansion or explosion ($t>0$).  The Coulomb explosion is asymptotically ballistic
\begin{equation}
\label{ballistic}
a(|t|\rightarrow \infty)=\sqrt{\frac{2E}{m}}|t|
\end{equation} 
which can be also seen directly from Eq.(\ref{energy_integral0}).

\item Eqs.(\ref{Hubble_definition})-(\ref{energy_integral0}) are Coulomb counterparts of the cosmological equations of the general theory of relativity which can be recovered via the Coulomb-Newton mapping correspondence relation
 \begin{equation}
\label{mapping}
e^{2}\rightarrow -Gm^{2}
\end{equation}
where $G$ is the universal gravitational constant.
\end{enumerate}

\subsection{Preview}

In what follows we will be studying dynamics of small perturbations away from the solutions described by Eqs. (\ref{Hubble_definition})-(\ref{energy_integral0}).  It is known that in hydrodynamics all possible small oscillations of the liquid about the state of rest or motion with constant velocity may be divided into oscillations of acoustic, vortical and entropic types with significant differences in character \cite{Monin,LL9}.  A similar division applies to the discussion of the oscillations about the Hubble flow, Eqs.(\ref{Hubble_law})-(\ref{energy_integral0}), as will become clear shortly. 

In our analysis we are guided by the accounts of the problem of gravitational instability \cite{ZN,Peebles,Mukhanov} modifying the reasoning as needed for the problem at hand.

\section{Static background}

Equations (\ref{continuity})-(\ref{general_Poisson}) have a trivial static solution $n=n_{0}$, $\textbf{v}=0$ that corresponds to the state of local neutrality.  This is a special $E=2\pi e^{2}(n_{0}\beta^{2})^{1/3}$ case of the general time-dependent solution (\ref{Hubble_law})-(\ref{energy_integral0}) when the Hubble parameter $H(t)$, Eq.(\ref{Hubble_law}), is identically zero.   Analysis of the dynamics of small disturbances away from this state is equivalent to the theory of sound that takes into account Coulomb interactions.  Even though no original conclusions will be reached, it is instructive to analyze this problem first with a greater degree of generality than is usually done \cite{Feynman}  as it exhibits many of the features of the case of interest when the background flows according to Eqs.(\ref{Hubble_law})-(\ref{energy_integral0}).  

\subsection{Linearized equations of motion}

Let us suppose that small perturbations of the density $\delta n$, velocity $\delta \textbf{v}$, potential $\delta \varphi$, pressure $\delta P$ and entropy $\delta s$ are superimposed onto the state of local neutrality $n=n_{0}$, $\textbf{v}=0$, $\varphi=0$, and $s=const$.  Substituting $n=n_{0}+\delta n$, $\textbf{v}=0+\delta \textbf{v}$,..., into Eqs.(\ref{continuity})-(\ref{general_Poisson}) and omitting the non-linear terms we find
\begin{equation}
\label{linearized_continuity}
\frac{\partial h }{\partial t}+\nabla\cdot\delta\textbf{v}=0,
\end{equation} 
\begin{equation}
\label{linearized_2nd_law}
\frac{\partial \delta\textbf{v}}{\partial t}=- \frac{e}{m}\nabla \delta\varphi-\frac{1}{mn_{0}}\nabla \delta P,
\end{equation}
\begin{equation}
\label{linearized_entropy_continuity}
\frac{\partial \delta s}{\partial t}=0,
\end{equation}
\begin{equation}
\label{linearized_Poisson}
\nabla^{2}{\delta\varphi}=-4\pi en_{0}h,
\end{equation}
where the dimensionless density contrast 
\begin{equation}
\label{constrast_definition}
h(\textbf{r},t)=\frac{\delta n(\textbf{r},t)}{n_{0}}
\end{equation}
is employed instead of $\delta n$ \cite{ZN,Peebles,Mukhanov}.  

\subsection{Entropic mode}

Eq.(\ref{linearized_2nd_law}) has a static solution 
\begin{equation}
\label{hydrostatic}
e\delta \varphi+\frac{\delta P}{n_{0}}=0
\end{equation}
expressing the condition of mechanical equilibrium of the Coulomb and pressure forces.  Corresponding density contrast $h(\textbf{r})$ can be found by employing the relationship
\begin{equation}
\label{differential}
\delta P= \left (\frac{\partial P}{\partial n}\right )_{s}\delta n+\left (\frac{\partial P}{\partial s}\right )_{n}\delta s
\equiv mc^{2}\delta n+\sigma\delta s,
\end{equation}  
where $c$ is the adiabatic speed of sound.  Combining the last two equations and eliminating the potential $\delta \varphi$ with the help of the Poisson equation (\ref{linearized_Poisson}) one finds 
\begin{equation}
\label{static_contrast}
\left (\omega_{p}^{2}-c_{0}^{2}\nabla^{2}\right )h=\frac{\sigma_{0}}{mn_{0}}\nabla^{2}\delta s, 
\end{equation}
where $\omega_{p}$ is the already mentioned plasma frequency (\ref{plasma_frequency}), and the subscript $0$ refers to the system's parameters evaluated at $n=n_{0}$.  Eq.(\ref{static_contrast}) implies that the source of the density contrast is the disturbance in the entropy $\delta s$.  Indeed,   Eq.(\ref{linearized_entropy_continuity}) has a static solution
\begin{equation}
\label{entropy_solution}
\delta s(\textbf{r},t)=\delta s(\textbf{r}),
\end{equation}
where $\delta s(\textbf{r})$ is the entropic perturbation at the moment of time $t=t_{i}$ when it was created.  Eq.(\ref{static_contrast}) can then be solved via a Fourier transform
\begin{equation}
\label{static_Fourier_solution}
h(\textbf{k})=-\frac{\sigma_{0}}{mn_{0}}\frac{k^{2}}{c_{0}^{2}k^{2}+\omega_{p}^{2}}\delta s(\textbf{k}),
\end{equation}
where $h(\textbf{k})$, the Fourier transform of $h(\textbf{r})$, is defined according to the convention
\begin{equation}
\label{Fourier_transform}
h(\textbf{k})=\int d^{3}r h(\textbf{r}) e^{-i\textbf{k}\cdot\textbf{r}},
\end{equation}
and similarly for other quantities of interest.  Inverting the Fourier transform in Eq.(\ref{static_Fourier_solution}) one finds
\begin{equation}
\label{static_contrast_solution}
h(\textbf{r})=-\frac{\sigma_{0}\delta s(\textbf{r})}{mn_{0}c_{0}^{2}}
+\frac{\sigma_{0}\omega_{p}^{2}}{4\pi mn_{0}c_{0}^{4}}\int dV'\frac{\delta s(\textbf{r}')}{|\textbf{r}-\textbf{r}'|}e^{-|\textbf{r}-\textbf{r}'|/d_{0}}
\end{equation}
where
\begin{equation}
\label{Debye}
d_{0}=\frac{c_{0}}{\omega_{p}}=\left (\frac{mc_{0}^{2}}{4\pi n_{0}e^{2}}\right )^{1/2}
\end{equation}
is the Debye screening length.  For perturbations whose spatial scale is significantly smaller than the Debye screening length $d_{0}$, Coulomb effects are negligible and $h(\textbf{r})$ approaches the first term in (\ref{static_contrast_solution}).  On the other hand, for sufficiently smooth perturbations Coulomb interactions dominate and $h(\textbf{r})=(\sigma_{0}/mn_{0}\omega_{p}^{2})\nabla^{2}\delta s(\textbf{r})$ as can be seen from Eq.(\ref{static_contrast}).  

Eqs.(\ref{entropy_solution})-(\ref{Debye}) summarize the properties of the entropic mode of the system.  Once created, it remains frozen in time;  if thermal conduction would be included, the entropic mode would become diffusive.

\subsection{Potential and vortical modes}

We now proceed to Eqs.(\ref{linearized_continuity}) and (\ref{linearized_2nd_law}) and divide the velocity disturbance into the potential (longitudinal) and vortical (transverse) parts $\delta \textbf{v}^{(l)}$ and $\delta \textbf{v}^{(t)}$ \cite{Monin,LL9} defined by
\begin{equation}
\label{velocity_split}
\delta\textbf{v}=\delta\textbf{v}^{(l)}+\delta\textbf{v}^{(t)},~~~~\nabla \times \delta\textbf{v}^{(l)}=0,~~~~\nabla\cdot \delta\textbf{v}^{(t)}=0
\end{equation}
Only the longitudinal part of the velocity disturbance enters the continuity equation (\ref{linearized_continuity})
\begin{equation}
\label{linearized_continuity_longitudinal}
\frac{\partial h}{\partial t}+\nabla\cdot\delta\textbf{v}^{(l)}=0
\end{equation} 
while the Euler equation (\ref{linearized_2nd_law}) separates into the two equations
\begin{equation}
\label{transverse_velocity}
\frac{\partial \delta\textbf{v}^{(t)}}{\partial t}=0,
\end{equation}
\begin{equation}
\label{linearized_2nd_law_longitudinal}
\frac{\partial \delta\textbf{v}^{(l)}}{\partial t}=- \frac{e}{m}\nabla \delta\varphi-c_{0}^{2}\nabla h-\frac{\sigma_{0}}{mn_{0}}\nabla\delta s,
\end{equation}
where we also employed Eq.(\ref{differential}).

\subsubsection{Vortical modes}

The equation for the vortical velocity (\ref{transverse_velocity}) is independent of the other equations;  its solution is
\begin{equation}
\label{transverse_solution}
\delta \textbf{v}^{(t)}(\textbf{r},t)=\delta \textbf{v}^{(t)}(\textbf{r}),
\end{equation} 
where $\delta \textbf{v}^{(t)}(\textbf{r})$ is the vortical perturbation at the moment of time $t=t_{i}$ when it was created.  Since $\nabla\cdot \delta \textbf{v}^{(t)}=0$, only two out of the three components of $\delta \textbf{v}^{(t)}(\textbf{r})$ are independent. They are vortical modes of the system.  A vortical perturbation does not perturb the density and once created remains frozen in time;  if shear viscosity was included, vortical modes would become diffusive. 

\subsubsection{Electro-acoustic modes}

If the operation of divergence is applied to both sides of Eq.(\ref{linearized_2nd_law_longitudinal}), one can employ the linearized continuity (\ref{linearized_continuity_longitudinal}) and the Poisson (\ref{linearized_Poisson}) equations to eliminate the potential $\delta \varphi$, resulting in the equation
\begin{equation}
\label{potential_eliminated}
\frac{\partial^{2}h}{\partial t^{2}}+(\omega_{p}^{2}-c_{0}^{2}\nabla^{2})h=\frac{\sigma_{0}}{mn_{0}}\nabla^{2}\delta s 
\end{equation}
which generalizes Eq.(\ref{static_contrast}).  In the long-wavelength limit (or if $P=0$), the $\nabla^{2}$ terms in (\ref{potential_eliminated}) can be omitted.  Then Eq.(\ref{potential_eliminated}) simplifies to the form
\begin{equation}
\label{electric_oscillation}
\frac{\partial^{2}h(\textbf{r},t)}{\partial t^{2}}+\omega_{p}^{2}h(\textbf{r},t)=0
\end{equation} 
without explicit dependence on the position $\textbf{r}$.  This describes simple harmonic motion with the plasma frequency $\omega_{p}$:
\begin{equation}
\label{general_electric_solution}
h(\textbf{r},t)=A(\textbf{r})\cos \omega_{p}t+B(\textbf{r})\sin\omega_{p}t.
\end{equation} 
The functions $A(\textbf{r})$ and $B(\textbf{r})$ can be determined given the density contrast $h(\textbf{r})$ and the velocity disturbance $\delta \textbf{v}(\textbf{r})$ at some initial moment of time $t=t_{i}$.  

Looking beyond the long-wavelength limit, Eq.(\ref{potential_eliminated}) can be turned into an ordinary differential equation via the Fourier transform: 
\begin{equation}
\label{general_solution_Fourier}
\frac{d^{2}h(\textbf{k},t)}{dt^{2}}+\omega^{2}(\textbf{k})h(\textbf{k},t)=-\frac{\sigma_{0}}{mn_{0}}k^{2}\delta s(\textbf{k}),
\end{equation}
where 
\begin{equation}
\label{spectrum}
\omega^{2}(\textbf{k})=\omega_{p}^{2}+c_{0}^{2}k^{2}
\end{equation}
determines the spectrum of the plasma waves \cite{Feynman}:  in the long-wavelength limit 
\begin{equation}
\label{lw_limit}
kd_{0}\ll 1
\end{equation}
one finds $\omega(\textbf{k})=\omega_{p}$ returning back to Eq.(\ref{electric_oscillation}) while in the acoustic limit $kd_{0}\gg 1$ one recovers the spectrum of the sound waves $\omega(\textbf{k})=c_{0}k$.  The general solution of the differential equation (\ref{general_solution_Fourier}) is the sum of the particular solution (\ref{static_Fourier_solution}) and 
\begin{equation}
\label{plasma_oscillation }
h(\textbf{k},t)=C(\textbf{k})\cos[\omega(\textbf{k})t]+D(\textbf{k})\sin[\omega(\textbf{k})t]
\end{equation} 
which describes the adiabatic $\delta s=0$ disturbance;  the functions $C(\textbf{k})$ and $D(\textbf{k})$ are determined by the initial conditions.  

The two independent solutions for the density contrast in Eqs.(\ref{general_electric_solution}) and (\ref{plasma_oscillation }) correspond to the two independent longitudinal electro-acoustic modes of the problem;  they are the focus of the classic treatments of small fluctuations in Jellium \cite{Feynman}.  

With the help of the mapping (\ref{mapping}) the analysis given above recovers a treatment of the gravitational instability of Einstein's static Universe \cite{Bonnor}.  Specifically, the spectrum (\ref{spectrum}) turns into $\omega^{2}(\textbf{k})=c_{0}^{2}k^{2}-4\pi Gmn_{0}$ with the implication that the system is unstable ($\omega^{2}(\textbf{k})<0$) with respect to long-wavelength fluctuations $k<k_{J}=\sqrt{4\pi Gmn_{0}}/c_{0}$.  Corresponding length scale $k_{J}^{-1}$, the Jeans length, is the gravitational counterpart of the Debye screening length (\ref{Debye}) of the Coulomb problem.    

\subsection{Summary}

To summarize, a generic small disturbance of Jellium away from the state of local neutrality $n=n_{0}$ can be presented in the form of a superposition of two vortical, two longitudinal electro-acoustic and one entropic perturbations which are the normal modes of the system.

\section{Hubble flow as a background}

Analysis of the dynamics of disturbances away from the state of local neutrality will be now used as a blueprint to investigate the evolution of small inhomogeneities imprinted on the Hubble flow described Eqs.(\ref{Hubble_law})-(\ref{energy_integral0}).  Substituting $n(\textbf{r},t)=n(t)+\delta n(\textbf{r},t)$, $\textbf{v}(\textbf{r},t)=H(t)\textbf{r}+\delta \textbf{v}(\textbf{r},t)$,..., into Eqs.(\ref{continuity})-(\ref{general_Poisson}), and omitting the non-linear terms we find
\begin{equation}
\label{Hubble_continuity}
\frac{\partial \delta n}{\partial t}+H(t)\textbf{r}\cdot\nabla \delta n+3H(t)\delta n +n(t)\nabla\cdot\delta \textbf{v} =0,
\end{equation}
\begin{equation}
\label{Hubble_Euler}
\frac{\partial \delta \textbf{v}}{\partial t}+H(t)(\textbf{r}\cdot\nabla)\delta \textbf{v}+H(t)\delta \textbf{v}=-\frac{e}{m}\nabla \delta \varphi-\frac{\nabla \delta P}{mn(t)},
\end{equation}
\begin{equation}
\label{Hubble_entropy}
\frac{\partial \delta s}{\partial t}+H(t)\textbf{r}\cdot\nabla \delta s=0,
\end{equation}
\begin{equation}
\label{Hubble_Poisson}
\nabla^{2}\delta \varphi=-4\pi en(t)h,
\end{equation}
where the density contrast $h$ is defined in a manner
\begin{equation}
\label{Hubble_contrast}
h(\textbf{r},t)=\frac{\delta n(\textbf{r},t)}{n(t)}
\end{equation}
that encompasses Eq.(\ref{constrast_definition}) as a special case.  Suggested by Eq.(\ref{comoving}), we seek a solution to the system of equations (\ref{Hubble_continuity})-(\ref{Hubble_Poisson}) that has the functional form 
\begin{equation}
\label{comoving_form}
h(\textbf{r},t)=h\left[\frac{\textbf{r}}{a(t)}, t \right]\equiv h(\textbf{x},t),
\end{equation}
and similarly for the remaining degrees of freedom of the problem.  This anticipates evolution of the background:  in the case of an expansion, $H(t)>0$, perturbations will be stretched out or "red-shifted" while for contraction, $H(t)<0$, perturbations will be compressed or "blue-shifted".  The ansatz (\ref{comoving_form}) is equivalent to a transformation into the reference frame comoving with the Hubble flow (\ref{comoving}):
\begin{equation}
\label{transformation}
\frac{\partial}{\partial t}+H(t) \textbf{r}\cdot\nabla=\left (\frac{\partial}{\partial t}\right )_{\textbf{x}}, ~~~\nabla=\frac{1}{a(t)}\nabla_{\textbf{x}},
\end{equation} 
where $(\partial/\partial t)_{\textbf{x}}$ stands for the partial time derivative for $\textbf{x}$ fixed while $\nabla_{\textbf{x}}$ refers to the vector differential operator with respect to the components of $\textbf{x}$.  With this in mind Eqs.(\ref{Hubble_continuity})-(\ref{Hubble_Poisson}) correspondingly transform into 
\begin{equation}
\label{Hubble_continuity_comoving}
\dot h+\frac{1}{a(t)}\nabla_{\textbf{x}}\cdot\delta\textbf{v}=0,
\end{equation}
\begin{equation}
\label{Hubble_Euler_comoving}
\dot{\delta \textbf{v}}+H(t) \delta \textbf{v}=-\frac{1}{a(t)}\left[\frac{e}{m}\nabla_{\textbf{x}}\delta\varphi+\frac{\nabla_{\textbf{x}}\delta P}{mn(t)}\right],
\end{equation}
\begin{equation}
\label{Hubble_entropy_comoving}
\dot{\delta s}=0,
\end{equation}
\begin{equation}
\label{Hubble_Poisson_comoving}
\frac{1}{a^{2}(t)}\nabla_{\textbf{x}}^{2}\delta \varphi=-4\pi e n(t)h,
\end{equation}
where the dot over dynamical variables is a shorthand for $(\partial/\partial t)_{\textbf{x}}$, and in arriving at Eq.(\ref{Hubble_continuity_comoving}) we employed Eqs.(\ref{Hubble_definition}) and (\ref{Hubble_contrast}).

It is instructive to compare Eqs.(\ref{Hubble_continuity_comoving})-(\ref{Hubble_Poisson_comoving}) with their $n=n_{0}$ counterparts, Eqs.(\ref{linearized_continuity})-(\ref{linearized_Poisson}).  Returning to the laboratory coordinates, $\nabla_{\textbf{x}}\rightarrow a(t)\nabla$,  would make the two settings more similar with time-dependent background density $n(t)$ replacing the constant density $n_{0}$ in Eqs.(\ref{Hubble_Euler_comoving}) and (\ref{Hubble_Poisson_comoving}).  Physically, the most significant difference can be seen when comparing corresponding Euler equations (\ref{linearized_2nd_law}) and (\ref{Hubble_Euler_comoving}) because the latter features an additional force of "Hubble friction" proportional to $-H(t)\delta \textbf{v}$.  Its effect on the evolution of inhomogeneities depends on the character of the Hubble flow.  Specifically, if the background is expanding, $H(t)>0$, the Hubble friction suppresses the growth of inhomogeneities while if it is contracting, $H(t)<0$, "Hubble anti-friction" operates and deviations away from uniformity grow.   

\subsection{Entropic mode}

The two sets of equations (\ref{Hubble_continuity_comoving})-(\ref{Hubble_Poisson_comoving}) and (\ref{linearized_continuity})-(\ref{linearized_Poisson}) are sufficiently different that a "comoving" counterpart of the condition of mechanical equilibrium (\ref{hydrostatic}) no longer exists.

On the other hand, Eq.(\ref{Hubble_entropy_comoving}) has a solution, a counterpart to Eq.(\ref{entropy_solution}), 
\begin{equation}
\label{comoving_entropy_solution}
\delta s(\textbf{x},t)=\delta s[a(t_{i})\textbf{x}]\equiv \delta s \left[\frac{a(t_{i})}{a(t)}\textbf{r}\right]
\end{equation}
that is static;  in the laboratory reference frame the entropic perturbation is advected by the Hubble flow.  Since the background flow for $n_{0}$ finite is an oscillation of the scale factor, the entropic perturbation (\ref{comoving_entropy_solution}) is then an oscillation of the same frequency.  On the other hand, if the underlying flow describes a Coulomb explosion, $n_{0}=0$, then asymptotically the scale factor diverges (\ref{ballistic}), and the entropic perturbation stretches to a constant without observable consequences.  Apart from different explicit $a(t)$ dependences, evolution of entropic perturbations described by Eq.(\ref{comoving_entropy_solution}) is the same as that found in  cosmology \cite{ZN,Peebles,Mukhanov}.

\subsection{Potential and vortical modes}

Since $\nabla \propto \nabla_{\textbf{x}}$, the velocity disturbance in Eqs.(\ref{Hubble_continuity_comoving}) and (\ref{Hubble_Euler_comoving}) can be again divided according to Eq.(\ref{velocity_split}) into the longitudinal  $\delta \textbf{v}^{(l)}$ and vortical $\delta \textbf{v}^{(t)}$ parts.  Only the former enters the continuity equation (\ref{Hubble_continuity_comoving})  
\begin{equation}
\label{Hubble_continuity_comoving_longitudinal }
\dot h+\frac{1}{a(t)}\nabla_{\textbf{x}}\cdot\delta\textbf{v}^{(l)}=0
\end{equation}
while the Euler equation (\ref{Hubble_Euler_comoving}) separates into the two equations
\begin{equation}
\label{transverse_velocity_comoving}
\dot {\delta\textbf{v}^{(t)}}+H(t)\delta \textbf{v}^{(t)}=0,
\end{equation}
\begin{eqnarray}
\label{linearized_2nd_law_longitudinal_comoving}
\dot {\delta\textbf{v}^{(l)}}&+&H(t)\delta \textbf{v}^{(l)}=\nonumber\\
&-&\frac{1}{a(t)} \left (\frac{e}{m}\nabla_{\textbf{x}} \delta\varphi+c^{2}\nabla_{\textbf{x}} h+\frac{\sigma}{mn}\nabla_{\textbf{x}}\delta s\right ),
\end{eqnarray}
where we also employed Eq.(\ref{differential});  the parameters of the equation of state $c^{2}$ and $\sigma$ are evaluated at $n=n(t)$, and are now time-dependent.

\subsubsection{Vortical modes}

Using the definition of the scale factor (\ref{scale_factor}) Eq.(\ref{transverse_velocity_comoving}) can be integrated with the result
\begin{equation}
\label{transverse_solution_comoving}
\delta \textbf{v}^{(t)}(\textbf{x},t)=\frac{a(t_{i})}{a(t)}\delta \textbf{v}^{(t)}[a(t_{i})\textbf{x}]\equiv \frac{a(t_{i})}{a(t)}\delta \textbf{v}^{(t)}\left[\frac{a(t_{i})}{a(t)}\textbf{r}\right]
\end{equation}
which means that as the vortical perturbation is advected by the flow, its amplitude is modulated inversely proportional to the scale factor.  In the presence of a charged background, $n_{0}\neq0$, the evolution of the vortical velocity (\ref{transverse_solution_comoving}) is an oscillation that has the same frequency as the background Hubble flow. 

On the other hand, in the case of the Coulomb explosion, $n_{0}= 0$, the scale factor diverges (\ref{ballistic}), and the vortical perturbation while stretching also falls off in magnitude as $1/t$.  

Apart from different explicit $a(t)$ dependences, the evolution of the two vortical perturbations described by Eq.(\ref{transverse_solution_comoving}) is the same as that found in cosmology \cite{ZN,Peebles,Mukhanov}.

\subsubsection{Electro-acoustic modes}

Subjecting both sides of the Euler equation (\ref{linearized_2nd_law_longitudinal_comoving}) to the operation of vector differentiation $\nabla_{\textbf{x}}$, and employing the continuity (\ref{Hubble_continuity_comoving_longitudinal }) and Poisson (\ref{Hubble_Poisson_comoving}) equations one arrives at the equation
\begin{equation}
\label{potential_eliminated_comoving}
\ddot{h}+2H(t)\dot{h}+\left (\frac{4\pi n e^{2}}{m}-\frac{c^{2}}{a^{2}}\nabla_{\textbf{x}}^{2}\right )h=\frac{\sigma}{mna^{2}}\nabla_{\textbf{x}}^{2}\delta s
\end{equation}
Apart from the Hubble friction $2H(t)\dot{h}$ term the structure of Eq.(\ref{potential_eliminated_comoving}) can be anticipated based on the appearance of its $n=n_{0}$ counterpart, Eq.(\ref{potential_eliminated}).  Indeed,  if in the expression for the square of the plasma frequency $\omega_{p}^{2}$ (\ref{static_contrast}) the constant background density $n=n_{0}$ is substituted by its time-dependent counterpart $n=n(t)$, similar replacements are made in the remaining entries of Eq.(\ref{potential_eliminated}), and the differential operator $\nabla$ is replaced with $\nabla_{\textbf{x}}/a$, Eq.(\ref{potential_eliminated_comoving}) would be largely recovered.  We also observe that in contrast to the case of the static background $n=n_{0}$, electro-acoustic and entropic modes are no longer decoupled.  Specifically, the entropic perturbation (\ref{comoving_entropy_solution}) that supplies the source term in the right-hand side of Eq.(\ref{potential_eliminated_comoving}) can generate the time-dependent density contrast $h$.  Below we limit our analysis to the two most practically relevant cases, when this effect is either negligible or strictly zero. 

(i)  In the long-wavelength limit (or if $P=0$) the $\nabla_{\textbf{x}}^{2}$ terms can be neglected, and Eq.(\ref{potential_eliminated_comoving}) simplifies to the form 
\begin{equation}
\label{electric_oscillation_comoving}
\ddot{h}+2H(t)\dot{h}+\frac{4\pi n(t) e^{2}}{m}h=0
\end{equation}      
without explicit dependence on $\textbf{x}$.  This is a generalization of Eq.(\ref{electric_oscillation}) to the case when the background flows according to Hubble's law (\ref{Hubble_law}).  Eq.(\ref{electric_oscillation_comoving}) has two linearly-independent solutions $h_{1,2}(t)$ so that its general solution can be written in the form
\begin{equation}
\label{general_electric_solution_comoving}
h(\textbf{x},t)=A(\textbf{x})h_{1}(t)+B(\textbf{x})h_{2}(t)
\end{equation}  
which is a counterpart to Eq.(\ref{general_electric_solution});  the functions $A(\textbf{x})$ and $B(\textbf{x})$ are determined by initial conditions.  The law of the evolution of the density and velocity perturbations determined by the functions $h_{1,2}(t)$ is however independent of their initial shape.   

If the Coulomb-Newton mapping correspondence relation (\ref{mapping}) is applied to Eq.(\ref{electric_oscillation_comoving}) one would recover the well-known cosmological equation $\ddot{h}+2H(t)\dot{h}-4\pi mGn(t)h=0$ \cite{ZN,Peebles,Mukhanov} that can be solved in closed form.  The same applies to Eq.(\ref{electric_oscillation_comoving}).  Specifically, one of its two independent solutions is
\begin{equation}
\label{Hubble_solution}
h_{1}(t)\propto H(t).
\end{equation}  
Indeed, differentiating both sides of Eq.(\ref{Friedmann1}) and combining the outcome with the definition of the Hubble parameter (\ref{Hubble_definition}) we obtain $\ddot{H}+2H(t)\dot{H}+\left[4\pi n(t)e^{2}/m\right]H=0$.  Comparing with Eq.(\ref{electric_oscillation_comoving}), we see that $h_{1}(t)\propto H(t)$ is one of its solutions.  Eq.(\ref{Hubble_solution}) has exactly the same appearance as its cosmological counterpart \cite{ZN,Peebles,Mukhanov}.

The second independent solution to Eq.(\ref{electric_oscillation_comoving}), $h_{2}(t)$, can be found with the help of the Wronskian 
\begin{equation}
\label{Wronskian}
W\equiv\dot{h_{1}}h_{2}-\dot{h_{2}}h_{1}=\frac{const}{a^{2}}.
\end{equation}
The right-hand side can be obtained by computing $\dot{W}$, combining the outcome with Eq.(\ref{electric_oscillation_comoving}), employing the definition of the scale factor (\ref{scale_factor}) followed by solution of the resulting differential equation $\dot{W}+2H(t)W=0$.  Substituting the ansatz $h_{2}=h_{1}f$ into the expression for the Wronskian (\ref{Wronskian}) one then obtains an equation for $f$ that can be integrated.  As a result, the second independent solution to Eq.(\ref{electric_oscillation_comoving}) is given by
\begin{equation}
\label{second_electric_solution}
h_{2}\propto H(t)\int^{t}\frac{dt}{a^{2}H^{2}}=H\int^{a}\frac{da}{\dot{a}^{3}}\propto H\frac{\partial t}{\partial E}
\end{equation}
where in arriving at the last representation we employed Eqs.(\ref{energy_integral0}) and (\ref{quadrature}).  The first two representations in Eq.(\ref{second_electric_solution}) are the same as in cosmology \cite{ZN,Peebles,Mukhanov}.  The last representation also implicit in the existing treatments  \cite{ZN,Peebles,Mukhanov} makes it possible to avoid the integration in Eq.(\ref{second_electric_solution}) and extract the solution $h_{2}$ from an explicit $a(t)$ dependence.

It is straightforward to verify that in the limit when the Hubble flow is a harmonic oscillation of the plasma frequency (\ref{quadrature}), the functions $h_{1,2}$ given by Eqs. (\ref{Hubble_solution}) and (\ref{second_electric_solution}) are simply the cosine and sine functions of $\omega_{p}t$.

Similarly, in the anti-de Sitter limit (\ref{anti}) one finds $h_{1}(t)\propto |\cot(\omega_{p}t/\sqrt{3})|$ and $h_{2}(t)\propto const$, both constrained by Eq.(\ref{constraint}).

The case when the background flow represents the Coulomb explosion can be visualized in terms of the $h_{1,2}$ dependences on the scale factor $a$ whose time dependence is in turn given by Eqs.(\ref{explosion_flow}).  Then the expression for $h_{1}$ (\ref{Hubble_solution}) follows from the $n_{0}=0$ limit of Eq.(\ref{energy_integral0}) while $\partial t/\partial E$ entering the expression for $h_{2}$ (\ref{second_electric_solution}) can be deduced from Eqs.(\ref{explosion_flow}).  As a result one finds
\begin{eqnarray}
\label{Coulomb_disturbances}
h_{1}&\propto& \frac{\sqrt{\alpha-1}}{\alpha^{3/2}},~~~~~~\alpha=\frac{a}{a_{1}}\nonumber\\
h_{2}&\propto& 1-\frac{3}{\alpha}-3 \frac{\sqrt{\alpha-1}}{\alpha^{3/2}}\ln(\sqrt{\alpha}-\sqrt{\alpha-1}) 
\end{eqnarray}   
where $a_{1}$, the scale factor corresponding to largest density of Jellium, is given by Eq.(\ref{constraint}).  

It is instructive to compare these expressions with their cosmological counterparts, $h_{1}\propto \alpha^{-3/2}\sqrt{1+\alpha}$, $h_{2}\propto 1+ 3/\alpha +3\alpha^{-3/2}\sqrt{1+\alpha}\ln(\sqrt{1+\alpha}-\sqrt{\alpha})$, describing the evolution of disturbances in an open, $E>0$, cosmological model without the cosmological constant \cite{Peebles}.  In both cases as $a\rightarrow \infty$ the first solution (\ref{Hubble_solution}) falls off as $h_{1}\propto 1/a\propto 1/t$ while the second solution (\ref{second_electric_solution}) saturates, $h_{2}\propto const$.  The latter conclusion is a consequence of the ballistic character of the late stage of the expansion (\ref{ballistic}).  Indeed, in this regime the Coulomb (or gravitational) forces become negligible and the perturbation ceases to evolve.       

\begin{figure}
\begin{center}
\includegraphics[width=1.0\columnwidth, keepaspectratio]{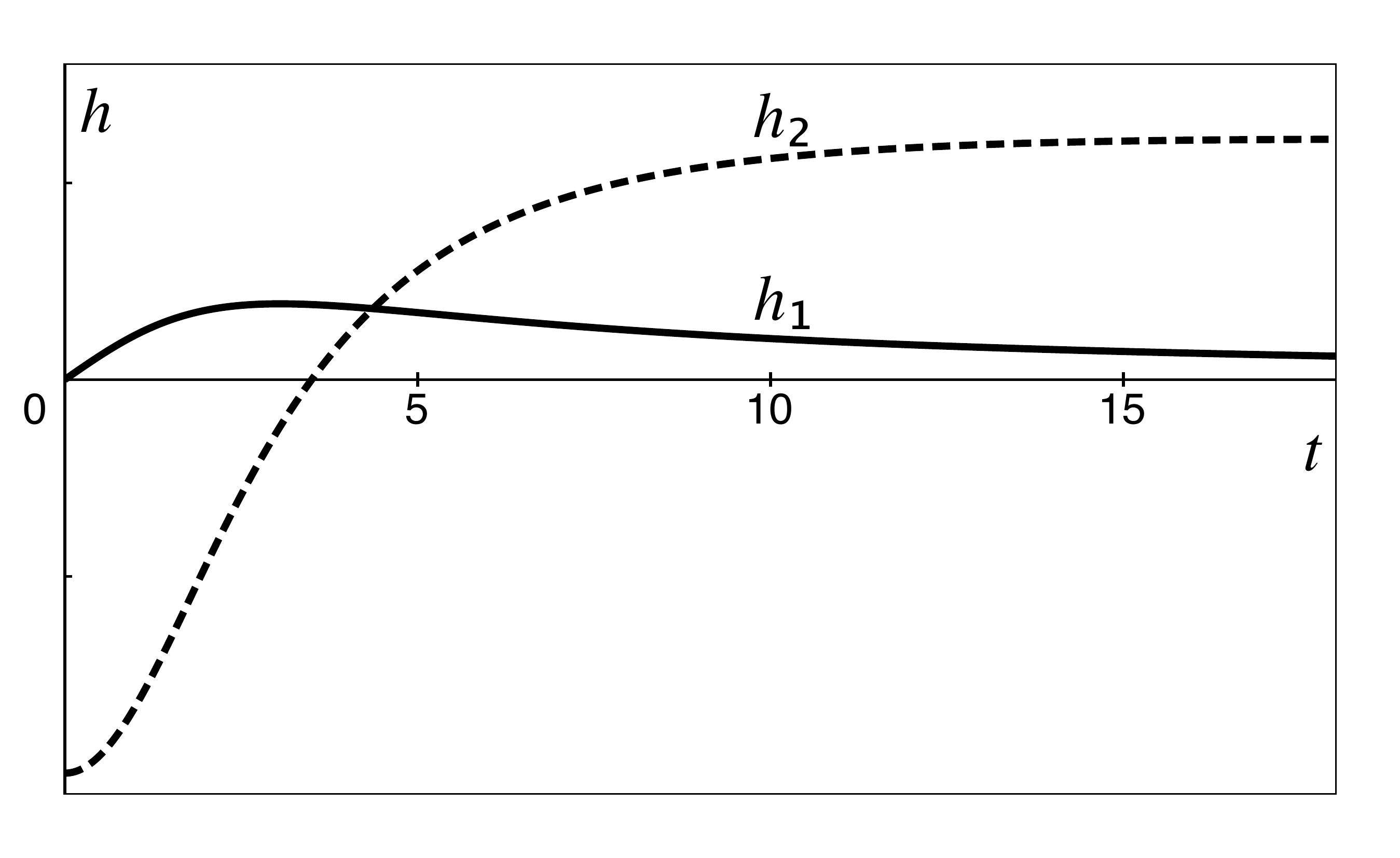} 
\caption{Evolution of the density contrast normal modes $h_{1,2}(t)$ (arbitrary units) according to Eqs.(\ref{Coulomb_disturbances}) and (\ref{explosion_flow}) for a Coulomb explosion with zero pressure equation of state or long-wavelength perturbations and arbitrary equation of state.  The unit of time is $(3ma_{1}^{3}/2\pi e^{2}\beta)^{1/2}/4$.}
\label{modes}
\end{center}
\end{figure}

Equations (\ref{Coulomb_disturbances}) and (\ref{explosion_flow}) determine the $h_{1,2}(t)$ dependences in parametric form;  they are shown in Figure \ref{modes}.
The first, asymptotically decaying mode $h_{1}(t)$ (bold curve), has a maximum at an intermediate time.  Although it is not easily discernible from Figure \ref{modes}, the same is true regarding the second, asymptotically saturating mode $h_{2}(t)$ (bold broken curve) so that the $h_{2}(t\rightarrow \infty)\rightarrow const$ limit is approached from above.  In this case the solution (\ref{second_electric_solution}) behaves as $h(\textbf{x},t\rightarrow \infty)\propto B(\textbf{x})=B(\sqrt{m/2E}\textbf{r}/t)$ where we employed Eq.(\ref{ballistic}).

(ii)  The entropic and electro-acoustic modes are also decoupled if the equation of state is isentropic, $P=P(n)$.  Then $\sigma=0$ and the right-hand side of Eq.(\ref{potential_eliminated_comoving}) vanishes.  The resulting partial differential equation can then be turned into an ordinary differential equation via a Fourier transform relative to the comoving coordinates $\textbf{x}$:
\begin{eqnarray}
\label{comoving_Fourier}
\frac{d^{2}h(\textbf{q},t)}{dt^{2}}&+&2H(t)h(\textbf{q},t)\nonumber\\
&+&\left[\frac{4\pi n(t)e^{2}}{m}+\frac{c^{2}(t)q^{2}}{a^{2}(t)}\right]h(\textbf{q},t)=0
\end{eqnarray}
where $h(\textbf{q},t)$, the Fourier transform of $h(\textbf{x},t)$, is defined according to the convention
\begin{equation}
\label{comoving_Fourier_definition }
h(\textbf{q},t)=\int d^{3}x h(\textbf{x},t)e^{-i\textbf{q}\cdot \textbf{x}}
\end{equation} 
Comparing with Eq.(\ref{Fourier_transform}) we see that a perturbation of a constant wave vector $\textbf{q}$ in the comoving reference frame (\ref{comoving}) corresponds to a perturbation of a time-dependent wave vector
\begin{equation}
\label{red_blue_shift}
\textbf{k}(t)=\frac{\textbf{q}}{a(t)}
\end{equation}
in the laboratory reference frame \cite{ZN,Peebles,Mukhanov}.

Eq.(\ref{comoving_Fourier}) is a second-order differential equation with variable coefficients which has two linearly-independent solutions $h_{1,2}(\textbf{q},t)$.  Its general solution can be written in the form
\begin{equation}
\label{comoving_plasma_oscillation}
h(\textbf{q},t)=C(\textbf{q})h_{1}(\textbf{q},t)+D(\textbf{q})h_{2}(\textbf{q},t)
\end{equation}
which is a counterpart to Eq.(\ref{plasma_oscillation });  the functions $C(\textbf{q})$ and $D(\textbf{q})$ are determined by initial conditions.  In the long-wavelength limit $q\rightarrow 0$ Eqs.(\ref{comoving_Fourier}) and (\ref{comoving_plasma_oscillation}) reduce to Eqs.(\ref{electric_oscillation_comoving}) and Eq.(\ref{general_electric_solution_comoving}).  The range of applicability of the latter can be written in a form that parallels the condition of the long-wavelength limit when the background is static (\ref{lw_limit}), $k(t)d(t)\ll1$, where $k(t)$ is given by Eq.(\ref{red_blue_shift}) while $d(t)$, the time-dependent counterpart of the Debye screening length, is given by substituting $c_{0}\rightarrow c(t)$ and $n_{0}\rightarrow n(t)$ in Eq.(\ref{Debye}). 

If the Coulomb-Newton mapping correspondence relation (\ref{mapping}) is applied to the differential equation (\ref{comoving_Fourier}), one would then recover the central result of the theory of  gravitational instability \cite{Bonnor}.  Specifically, the expression in the square parentheses would become $c^{2}(t)q^{2}/a^{2}(t)-4\pi Gmn(t)$, and if it would be negative, the mode of the wave vector $\textbf{q}$ would be unstable.  

Similar reasoning applied to Eq.(\ref{comoving_Fourier}) rules out any instability because the expression in the square parentheses is always non-negative.  Specifically, combining with our earlier conclusions regarding the entropic and vortical modes, we conclude that the Coulomb explosion is stable with respect to \textit{all} possible kinds of small perturbations.  Our results disprove the claim \cite{Kaplan} that any perturbation away from Hubble expansion would lead to formation of shock waves.      

In order to gain more insight into the character of its solutions, it is instructive to rewrite Eq.(\ref{comoving_Fourier}) in the form of a Schr\"odinger-type equation
\begin{equation}
\label{Schrodinger}
\frac{d^{2}}{dt^{2}}\left[h(\textbf{q},t)a(t)\right]+\omega^{2}(\textbf{q},t)\left[h(\textbf{q},t)a(t)\right]=0,
\end{equation}      
where 
\begin{equation}
\label{time-dependent-spectrum}
\omega^{2}(\textbf{q},t)=\frac{c^{2}(t)q^{2}}{a^{2}(t)}+\frac{8\pi e^{2}}{3m}\left[n(t)+\frac{n_{0}}{2}\right]
\end{equation}
is the time-dependent counterpart of the dispersion law (\ref{spectrum}).  

If the function $\omega(\textbf{q},t)$ varies adiabatically, $\dot{\omega}(\textbf{q},t)\ll \omega^{2}(\textbf{q},t)$, the semi-classical approximation holds and the closed form solutions to Eq.(\ref{Schrodinger}) could be given as \cite{LL3}
\begin{equation}
\label{WKB_solution}
h_{1,2}\propto \frac{1}{a(t)\sqrt{\omega(\textbf{q},t)}}\exp\left[\pm i\int^{t}\omega(\textbf{q},t')dt'\right]
\end{equation}
In the short-wavelength limit when the first term in Eq.(\ref{time-dependent-spectrum}) dominates, the solution (\ref{WKB_solution}) simplifies to the form 
\begin{equation}
\label{WKB_solution_short_waves}
h_{1,2}\propto \frac{1}{\sqrt{c(t)a(t)}}\exp\left[\pm iq\int^{t}\frac{c(t')dt'}{a(t')}\right]
\end{equation}
already given in cosmology \cite{Mukhanov}.  

Eq.(\ref{WKB_solution_short_waves}) also covers the asymptotic $t\rightarrow \infty$ limit of the Coulomb explosion problem.  For example, for the $c(t)=const$ equation of state Eqs.(\ref{WKB_solution_short_waves}) and (\ref{ballistic}) predict that $h_{1,2}(t\rightarrow \infty)\propto t^{-1/2\pm iqc\sqrt{m/2E}}$. 

\section{Conclusions}

To summarize,  generic small disturbance around Hubble flows in Jellium can be resolved into a superposition of two electro-acoustic, two vortical, and one entropic perturbations which are the normal modes of vibration of the system.    One of the distinctive features of the electro-acoustic modes is that they can only have a form of standing waves;  traveling wave solutions do not seem possible.

It is the present author's hope that the normal mode analysis carried out in this work will stimulate experimental and numerical studies of Jellium-type systems.  On the experimental side, Hubble flows in conductors may be excited by passing an electron beam through a metal nanosphere and then inferring existence of  the flow via electron energy loss spectroscopy, a technology employed to detect plasmons.  Indeed, there exists an experimental evidence that breathing modes that resemble Hubble flows can be excited this way in Silver nanodisks \cite{silver}.  In Coulomb explosion setting relevant experiments should involve extended plasmas.     

\section{Acknowledgements}

I would like to thank E. Sarabamoun for careful reading of the manuscript.

\end{document}